\documentstyle[prl,aps,multicol,epsfig]{revtex}
\begin{document}
\title{Magneto-electrodynamics at high frequencies in the antiferromagnetic and superconducting states of DyNi${}_2$B${}_2$C}
\author{Durga P. Choudhury${}^{1,2}$, H. Srikanth${}^1$ and S. Sridhar${}^1$}
\address{${}^1$Physics Department, Northeastern University, Boston, MA 02115}
\address{${}^2$Rome Laboratory, Hanscom Air Force Base, Bedford, MA 01730}
\author{P. C. Canfield}
\address{Ames Laboratory, Department of Physics and Astronomy, Iowa State University, IA 50011}
\draft
\date{\today}
\maketitle
\newcommand{\T}{\textstyle}
\begin{abstract}

We report the observation of novel behaviour in the radio frequency ({\it rf\/})
and microwave response of $\rm DyNi_2B_2C$ over a wide range of
temperature ($T$) and magnetic field ($H$) in the antiferromagnetic (AFM)
and superconducting (SC) states. At microwave frequencies of 10\thinspace
GHz, the $T$ dependence of the surface impedance $Z_s=R_s+iX_s$ was
measured which yields the $T$ dependence of the complex conductivity
$\sigma_1-i\sigma_2$ in the SC and AFM states. At radio frequencies
(4\thinspace MHz), the $H$ and $T$ dependence of the penetration depth
$\lambda(T,H)$ were measured.

The establishment of antiferromagnetic order at $T_N=10.3$\thinspace $K$
results in a marked decrease in the scattering of charge carriers, leading
to sharp decreases in $R_s$ and $X_s$. However, $R_s$ and $X_s$
differ from each other in the AFM state. We show that the results are
consistent with conductivity relaxation whence the scattering rate becomes
comparable to the microwave frequency. 

The {\it rf\/} measurements yield a rich dependence of the scattering on the
magnetic field near and below $T_N$. Anomalous decrease of scattering at
moderate applied fields is observed at temperatures near and above $T_N$,
and arises due to a crossover from a negative magnetoresistance state, possibly
associated with a loss of spin disorder scattering at low fields, to a positive
magnetoresistance state associated with the metallic nature. The normal state
magnetoresistance
is positive at all temperatures for $\mu_0H>2T$ and at all fields for
$T>15K$. Several characteristic field and temperature scales associated with
metamagnetic transitions ($H_{M1}(T)$, $H_{M2}(T)$) and onset of spin
disorder $H_D(T)$, in addition to $T_c$, $T_N$ and $H_{c2}(T)$ are
observed in the {\it rf\/} measurements.

\end{abstract}
\begin{multicols}{2}
\section{Introduction}

The discovery of superconductivity in the quaternary borocarbides\cite
{RNagarajan94a,RJCava94a} has led to a new family of materials with
intriguing properties\cite{KDDRathnayaka96a,YDeWilde97a}. These are the only
known intermetallic
superconductors containing an appreciable amount of Nickel that have
transition temperatures in excess of 10\,K. Stoichiometric single phase
compounds with the general composition of $R {\rm Ni_2B_2C}$, where $R$ is a
rare earth element, Yttrium or Lutetium have been synthesized\cite{RJCava94b}
and extensively investigated. Most members of this family that exhibit
superconductivity also have coexisting antiferromagnetic order, notable
exceptions being ${\rm YNi_2B_2C}$ and ${\rm LuNi_2B_2C}$\cite{HEisaki94a}.
This system therefore is very useful in studying the interplay between
magnetism and superconductivity. The order in temperature in which these
two transitions occur
as well as the values of the relevant temperature scales associated ($T_N$
for antiferromagnetic transition and $T_c$ for superconducting transition)
vary significantly. Although the monotonic decrease in superconducting
transition temperature with increasing de\,Gennes factor of the rare earth is
usually attributed to the pair breaking effects of the moments\cite{BKCho95a},
the magnetism in these
materials is much more complex and defies a simple explanation. For one
thing, it turns out that the presence of Nickel, hitherto thought to be
grossly detrimental to superconductivity because of it's pair breaking
effects, is relatively benign in these materials and does not contribute
significantly to it's magnetic properties\cite{MEMassalami95a}. Most of the
pair breaking comes from the rare earth moments in these systems where they
are present.

Recently, superconductivity ($T_c=6.2$\,K) was reported in ${\rm DyNi_2B_2C}$,
below an antiferromagnetic transition at $T_N=10.3$\,K\cite
{BKCho95a,CVTomy95b}. Thus this material is one of a select few materials
where superconductivity appears in the presence of long-range magnetic
order, some others being systems like ${\rm Ho(Ir_xRh_{1-x})_4B_4}$
\cite{HCKu80a}, ${\rm Tb_2Mo_3Si_4}$\cite{FGAliev94a} and some of the
heavy fermion superconductors.

Metallic or superconducting antiferromagnets afford the possibility to
study the influence of local moments on electronic properties. In the
metallic state, interesting effects are expected on the transport properties
such as the conductivity. In the superconducting state, magnetism can lead
to novel effects on the superconducting order parameter. This is because of
the presence of low energy and magnetic field scales in the system, such as
spin-flop field scales arising from the magnetic state, and vortex-related
fields due to superconductivity.

High frequency measurements such as those of the surface impedance
$Z_s=R_s+iX_s$ yield unique information often not available with other
techniques, particularly those at dc or low frequencies such as dc
resistivity or ac susceptibility. In the metallic state, information
regarding the dynamics of electrons and electronic moments can be obtained.
In the superconducting state, the order parameter can be probed in terms
of the penetration depth. Because of finite dissipation at high frequencies,
one can simultaneously probe the quasiparticle effects too.

In this paper we report our results on measurements of radio frequency
($\rm \sim 4\,MHz$) skin depth and microwave ($\rm \sim 10\,GHz)$ surface
impedance on high quality single crystals of this material. Several novel
features are reported. A marked decrease in the scattering of charge carriers
is observed upon establishment of antiferromagnetic order at
$T_N=10.3$\thinspace $K$. $R_s$ and $X_s$ differ from each other in the
antiferromagnetic state, which we show arises from conductivity relaxation.
Anomalous decrease of scattering at moderate applied fields is observed at
temperatures near and above $T_N$, and is proposed to arise from changes
in spin disorder scattering. Several characteristic field and temperature
scales associated
with metamagnetic transitions ($H_{M1}(T)$, $H_{M2}(T)$) and a crossover
field scale $H_D(T)$, in addition to $T_c$, $T_N$ and $H_{c2}(T)$ are
observed in the {\it rf\/} measurements.The behaviour of the complex
conductivity in the SC state and the AFM state are obtained as functions of
temperature.

\section{Experimental Details}

\subsection{RF Setup}

The {\it rf\/} measurements were carried out in a tunnel diode driven tank
oscillator self-resonant typically at 4\,MHz. The sample is placed in an
inductive coil which is part of the L-C tank with the {\it rf\/} field $H_{rf}$,
the dc field $H_{dc}$ and the \^c axis oriented such that $H_{rf}\perp
H_{dc}\perp$\,\^c. The inset to Fig.\ref{Fig02} shows a picture of this
geometry. Magnetic fields upto 70 kOe were applied using a superconducting
magnet. This experimental technique has been extensively used to study high
and low $T_c$ superconductors both in the Meissner and mixed states\cite
{DHWu90a,SOxx96a,SSridhar96a}. If either the skin depth $\delta$ or the
superconducting penetration depth $\lambda$ changes as functions of $T$ or
$H$, these changes can be measured as changes $\Delta f$ in resonant
frequency using $\Delta(\delta \,{\rm or}\,\lambda )=-G\,\,\Delta f$, where
$G$ is a geometric factor. The oscillator is ultra-stable, (approximately
1\,Hz in 4\,MHz), and this leads to very high sensitivity, with typical
resolutions of a few \AA . Earlier experiments in cuprate superconductors
have led to a wide variety of information regarding superconducting
parameters such as the lower critical field $H_{c1}$ and pinning forces
$\kappa_P$\cite{DHWu90a} and have also been recently used to study
borocarbide superconductors where they have revealed new features in the
$H$-$T$ phase diagram\cite{SOxx96a,SSridhar96a}. The primary advantage of our
technique is the ability to probe both the resistivity in the normal state
and superfluid density in the superconducting state which is not possible
with conventional techniques such as DC resistivity or AC susceptibility. An
interesting example of the versatility and sensitivity of the setup was
recently demonstrated when we could resolve an {\it area preserving}
hexagonal to square transition of the flux line lattice in ${\rm ErNi_2B_2C}$%
\cite{MREskildsen97a}, another member of the borocarbide family, which was not
visible in magnetic susceptibility measurements.

The sample was mounted on a sapphire rod with a groove machined on it.
Sapphire is a good thermal conductor but a poor electrical conductor; thus
while the sample is thermally well anchored to it's environment, the
contribution of the sample holder to the frequency shift is minimal. The
coil was mounted on the sapphire and was attached to a sheet of alumina with
GE varnish. The alumina sheet carried an electric heater for fine tuning of
the sample temperature. Data acquisition was done by a computer using a
GPIB interface.

\subsection{Microwave Setup}

The microwave measurements were carried out in a superconducting Niobium
cavity using the ``cold cavity-hot finger'' technique\cite{SSridhar88b}. The
cavity is immersed in a bath of liquid Helium, and the sample is mounted on
a piece of sapphire inside the cavity. The sapphire has a heater mounted on
it so that the temperature of the sample can be varied from below 4\,K to
above 200\,K while the cavity is maintained at a fixed temperature of
4.2\,K, as is necessary for it to remain superconducting. The perturbation
of the cavity due to the change in the superconducting properties of the
sample is reflected as a change of resonance frequency $f$ and bandwidth
$\delta f$ of the cavity. This is related to the change in the complex
surface impedance ($Z_s = R_s+iX_s$) as $R_s=\Gamma(1/Q_s(T)-1/Q_0(T))$ and
$\Delta X_s=2\pi\mu_0f_0\Delta\lambda=(-2\Gamma/f_0) (f_s(T)-f_0(T))$ where
$\Gamma$ is a geometric factor relating to the cavity and the sample
dimensions, $Q_s$ and $Q_0$ are the loaded and unloaded quality factor of
the cavity ($Q=f/\delta f$), and $f_s$ and $f_0$ are the loaded and unloaded
resonance frequencies respectively. Our setup allows us to measure the
absolute value of $R_s$ but only the relative change in $X_s$. An absolute
value is imposed on $X_s$ by assuming that $X_s=R_s$ at high temperatures ($%
T\gg T_c,\,T_N$) where any effect of superconductivity or magnetism would be
negligible. The superconducting cavity has very high Q, of the order of $%
10^7 $-$10^8$, making it very sensitive to small changes in the surface
resistance ($\sim 10\mu\Omega$) and penetration depth ($\sim 1$\AA). An
additional advantage of the superconducting cavity is that it shields out
stray magnetic fields from the sample. This technique has been very
successfully used to yield a wealth of information on other members of the
borocarbide family\cite{TJacobs95b} as well as many high temperature
superconductors\cite{TJacobs95d,HSrikanth97a}

\subsection{Crystal Growth}

High quality single crystals of the material were prepared using the high
temperature flux growth technique\cite{MXu94a}. This method consists of
furnace heating a mass of stoichiometric polycrystalline material with an
equal mass of ${\rm Ni_2B}$ flux in an inert atmosphere. Single crystals of
the material grow into the flux in the shape of platelets. The crystals grown
this way have relatively large size, making
them particularly useful for measurement of bulk properties. The superiority
of these crystals in terms of phase purity, proper stoichiometry and low
density of defects have been verified by X-ray diffraction and other methods%
\cite{BKCho95a}. Typical size of the crystal used in our experiments is $%
2\times 0.8\times 0.2\,{\rm mm}^3$.

\section{Results and Discussion}

\subsection{Radio Frequency Measurements}

These experiments, carried out for ${\rm 2\,K}<T<{\rm 100\,K}$ and $0<H<70\,{\rm kOe}$
essentially measure the real part of the complex electromagnetic
penetration depth $\tilde\lambda=\sqrt{\frac{\textstyle -i}{\textstyle\mu\omega\sigma}}$
at radio frequencies. In the normal state, $\sigma=\sigma_n$ is the normal
conductivity and is purely real. In this
case, $\tilde\lambda^{-1}=\delta^{-1}(1+i)$, where
$\delta=\sqrt{\frac{\textstyle 2}{\textstyle \mu\omega\sigma_n}}$ is the skin depth. In the
superconducting state, $\sigma$ has to be replaced with an effective complex
conductivity $\sigma_s=\sigma_1-i\sigma_2$ whose real and imaginary parts
are proportional to the quasiparticle scattering and superfluid density
respectively. In the limit $\sigma_2\gg\sigma_1$, which typically holds for
$T<T_c$, $\tilde\lambda={\frac{\textstyle 1}{\sqrt{\textstyle\mu\omega\sigma_2}}}=\lambda_L$,
the London penetration depth. Notice that the normal
state skin depth is a factor of $\sqrt 2$ bigger than the penetration depth.
All the data presented in this paper show penetration depths both in the
normal and superconducting states and are denoted by $\lambda$.

The temperature dependence of $\lambda$ shows a clear signature of the onset
of the magnetic order at 10.3\,K, (see fig.\ref{Fig01}) which has also been observed by other
techniques such as neutron scattering and specific heat measurements by
other investigators\cite{PDervenagas95a,MSLin95a,JWLynn96b}. Onset of the
superconducting transition brings about
further decrease in magnetic field penetration. Note that measurement of DC
resistivity also shows a very similar behaviour\cite{BKCho95a} which is to
be expected because of the proportionality between the square root of
resistivity and the skin depth as mentioned above. Although no details of
temperature dependence of penetration depth in the superconducting state is
visible in DC measurements because of the very nature of the technique, in
our setup this information is clearly seen.
{
\narrowtext
\begin{center}
\begin{figure}
\mbox{\epsfig{file=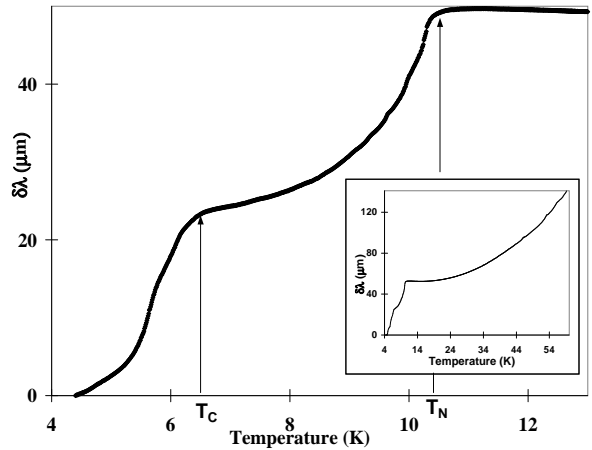, height=3.2 truein, width=2.5 truein, angle=-90}}
\caption{Electromagnetic screening depth $\lambda$ against temperature. The
	onset of superconductivity and the signature of the antiferromagnetic order
	are quite visible. Notice the similarity with fig.\,3(a) of ref.\,7
	which is expected because of the proportionality between $\lambda$ and
	$\sqrt\rho$. Inset: The same data shown over extended temperature range.}
\label{Fig01}
\end{figure}
\end{center}
}
The same experiment, done in the presence of a finite DC field, reveals
further interesting features. Fig.\ref{Fig02} shows typical data taken for
$H=0$, 2.5\,kOe, 5\,kOe and 10\,kOe respectively. The
inset show the relative orientation of the sample's \^c-axis and the applied
dc and {\it rf\/} fields. The direction of the induced {\it rf\/} current is
also shown as a reference. Successive curves are shifted along the vertical
axis by roughly the same amount to render greater clarity to the figure.
Since in this experiment we only measure the {\it change\/} in screening
length, a constant shift will not affect any of the further discussion. The
total change in screening length is shown on the right hand side of each
curve. The superconducting transition is still visible but with a depressed
$T_c$ in the field-cooled $\lambda(T)$ curve at $H=2.5$\,kOe. The application
of a magnetic field decreases the sharpness of the change of $\lambda$ at
the antiferromagnetic transition and it appears that at a field greater than
the zero temperature $H_{c2}$, the $\lambda$ would attain a constant value
independent of temperature at very low temperatures. This reflects the
residual resistivity of the material. As temperature is increased, a dip
appears in the screening length for field scales $\sim 10$\,kOe. It signals
the existence of some mechanism that leads to the reduction in scattering of
charge carriers. We shall elaborate on that later in this paper.
{
\narrowtext
\begin{center}
\begin{figure}
\mbox{\epsfig{file=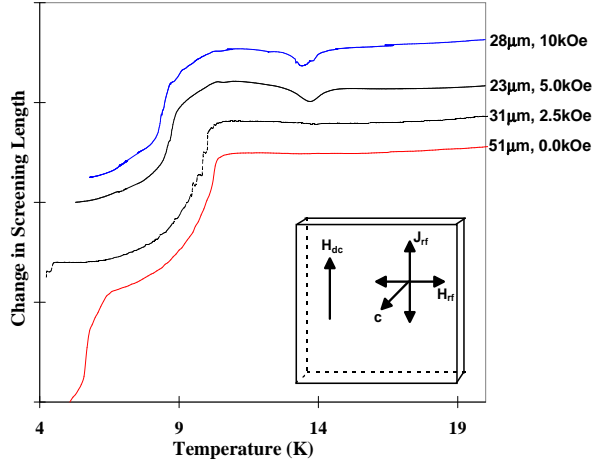, height=3.2 truein, width=2.5 truein, angle=-90}}
\caption{Radio frequency penetration depth against temperature measured in
	finite magnetic field. Various data sets have been shifted from each other
	vertically for clarity of presentation. Inset : Schematic of sample
	orientation relative to $H_{dc}$ and $H_{rf}$.}
\label{Fig02}
\end{figure}
\end{center}
}
To see these features from another perspective, we did a series of
measurements of $\lambda$ against $H$ at various fixed temperatures and a
few typical data sets are shown in Fig.\ref{Fig03}. The plots for
successively increasing temperatures have been displaced from each other
along the vertical axis to avoid overlapping, as was done in Fig.\ref{Fig02}.
Several features of the field ramp data, although present in the temperature
ramps, are now more obvious. We point them out in the following.

\begin{itemize}
\item  The magnetic state shows two sharp jumps of the screening length both
above and below $T_{c}$, originating from metamagnetic transitions in the
sample. Such transitions have also been noticed in DC magnetization
measurements of ${\rm DyNi_{2}B_{2}C}$\cite{MSLin95a,PCCanfield97b} as well as in other
members of the borocarbide family such as ${\rm HoNi_{2}B_{2}C}$\cite
{SOxx97a,PCCanfield94a,PCCanfield97a}. With the increase of temperature, the magnitude of
the jump at field $H_{M1}$ (as defined in Fig.\ref{Fig04}) decreases and that
of the one at $H_{M2}$ increases. As a result of these transitions, the mean
free path in the sample decreases rapidly as field is increased from zero to
about 10\thinspace kOe.

\item  In the non-superconducting state, regardless of the magnetic order,
the resistance
goes through a local minimum as the field is swept from zero. The sharpness of
this minimum goes down as temperature is increased, and at the highest
temperature of 17.7K at which data was taken, this minimum is barely
visible. We remark here in passing that the data shown in Fig.\ref{Fig02}
indicates that the field scale $H_{D}$ (defined in Fig.\ref{Fig04}) at which
this dip is most pronounced decreases rapidly at $T\sim 13$\thinspace K.

\item  The response of the material at very high field (${\rm >20kOe}$)
shows positive magnetoresistance at all temperatures, characteristic of it's
metallic behaviour. Data is shown only upto 40\,kOe in fig.\ref{Fig03} for
clarity of presentation. Beyond this, $\lambda$ increases monotonically with
field up to 70\,kOe, the maximum field at which data was taken.
\end{itemize}
{
\narrowtext
\begin{center}
\begin{figure}
\mbox{\epsfig{file=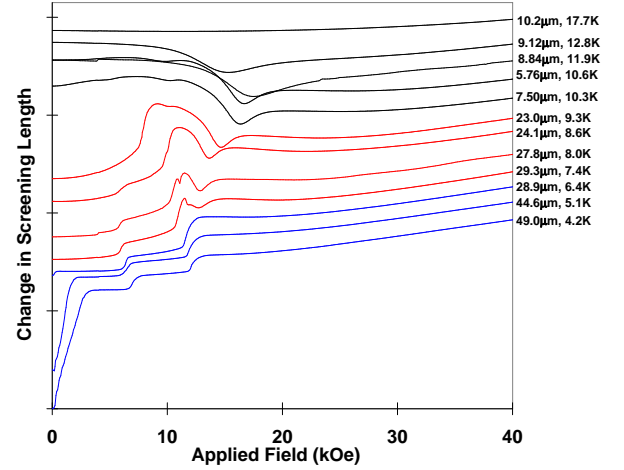, height=3.2 truein, width=2.5 truein, angle=-90}}
\caption{Screening length against field at various temperatures. Each curve
	has been displaced along the vertical axis by a fixed amount for better
	visibility. The total change in screening length [$\lambda(H={\rm 70 kOe},T)
	-\lambda(H=0,T)$] is indicated on the right hand side for each curve.}
\label{Fig03}
\end{figure}
\end{center}
}
It is obvious from Fig.\ref{Fig03} that the screening length is characterized
by field scales in addition to the well known scale of $H_{c2}$. In fig.\ref
{Fig04} we define the nomenclature for these field scales that we will
subsequently be using for discussion throughout this paper. The two field
scales corresponding to the metamagnetic transitions are called $H_{M1}$ and 
$H_{M2}$ respectively and the scale at which skin depth reaches a local minimum
for $T>T_c(H=0)$ is called $H_D$.
{
\narrowtext
\begin{center}
\begin{figure}
\mbox{\epsfig{file=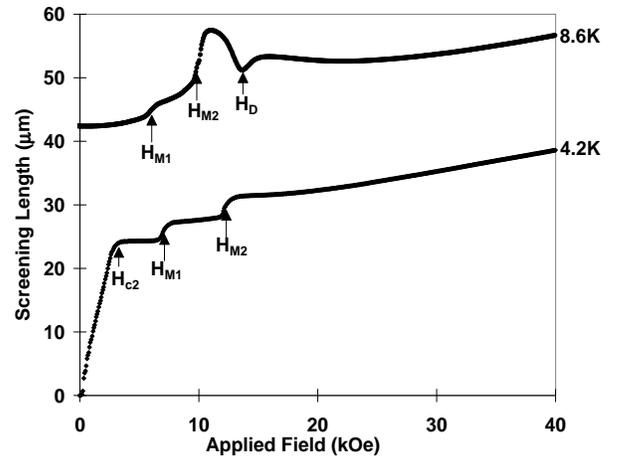, height=3.2 truein, width=2.5 truein, angle=-90}}
\caption{Characteristic field scales in DyNi${}_2$B${}_2$C.}
\label{Fig04}
\end{figure}
\end{center}
}
To further analyze these features, we numerically calculated $d\lambda/dH$
for the data shown in Fig.\ref{Fig03}. Three typical data sets of this kind,
one each for the case of $T<T_c$, $T_c<T<T_N$ and $T>T_N$ are shown in
fig.\ref{Fig05}. The characteristic field scales that we discussed above are now
much more obvious.

The topic of the interplay between antiferromagnetism and superconductivity
has been extensively investigated\cite
{AIBuzdin86a,MBMaple95a,RKonno95a,MBMaple82,MLKulic95a}. Many of the
``classic'' antiferromagnetic superconductors like the Rhodium Boride
cluster compounds and $R{\rm Mo}_8{\rm S}_8$ ($R$ = Rare Earth) are well
known to undergo field induced spin-flop transitions in the superconducting
state which have been investigated both experimentally\cite{HIwasaki86a} and
theoretically \cite{AIBuzdin90a,OWong89a}. Such transitions in
${\rm DyNi}_2{\rm B}_2{\rm C}$ occur at $H_{M1}$ and $H_{M2}$ ($>H_{c2}$) after superconductivity has
been quenched, either by raising temperature or the applied field.

Above $T_N$, the magnetoresistance of this material goes through a local minimum
with increasing field (Fig.\ref{Fig03}), the sharpness of
which decreases as temperature increases.
Although at first inspection this local minimum in magnetoresistance may appear
quite novel, we believe that it is associated with a crossover from the negative
magnetoresistance associated with the loss of spin disorder scattering from the
Dy sublattice as it is aligned along the applied field to the the positive
magnetoresistance associated with conduction band electrons. These two effects
are clearly seen in data taken in the paramagnetic state of $\rm HoNi_2B_2C$ and
$\rm LuNi_2B_2C$\cite{IRFisher96a,IRFisher97a}. Although these two effects do
indeed give rise to a local minimum in magnetoresistance in the paramagnetic state
they do not give rise to as sharp a local minimum as is seen in the data in
fig.\ref{Fig03}. This enhanced sharpness may be associated with interaction between
the local moments since $T\approx T_N$ for these data. Another possible explanation
for the local minimum is that it represents an as-of-yet-undetected phase transition
line that exists at intermediate fields for a temperature near $\rm T_N$. Although
we currently consider this to be the less likely explanation, the possibility of
another phase transition is the subject of ongoing research.
{
\narrowtext
\begin{center}
\begin{figure}
\mbox{\epsfig{file=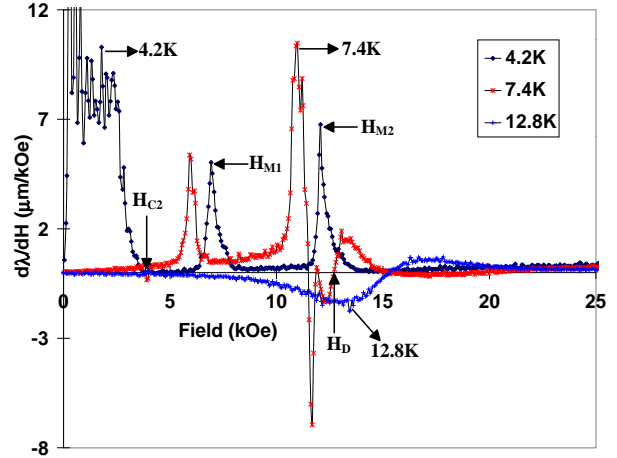, height=3.2 truein, width=2.5 truein, angle=-90}}
\caption{$d\lambda/dH$ for DyNi${}_2$B${}_2$C at different temperatures.}
\label{Fig05}
\end{figure}
\end{center}
}
From the data of Fig.\ref{Fig05}, we constructed a ``phase diagram'' for
DyNi${}_2$B${}_2$C demarcating the temperature dependence of the various field
scales as shown in Fig.\ref{Fig06}. $H_{c2}$ was defined to be the field at
which $d\lambda/dH$ becomes vanishingly small. The values thus obtained
are somewhat higher than those obtained from DC measurements\cite{BKCho95a}.
The same criterion was used for determining $H_D$, the field
at which $\lambda$ goes through a minimum. It is worth mentioning here that
$H_D$ most likely a crossover field scale and not does not represent a phase
transition as was discussed previously. $H_{M1}$ and $H_{M2}$, however,
represent field scales corresponding to metamagnetic transitions\cite
{ZHossain96a} and are chosen so that $d\lambda/dH$ is maximum. The region
within the full width at half maximum of each peak is shaded in the diagram
demarcating the approximate span of the transition region.
{
\narrowtext
\begin{center}
\begin{figure}
\mbox{\epsfig{file=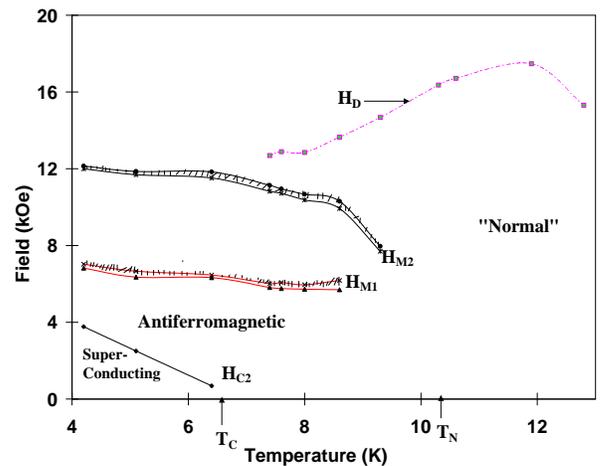, height=3.2 truein, width=2.5 truein, angle=-90}}
\caption{Temperature dependence of characteristic field scales in
	DyNi${}_2$B${}_2$C. The shaded regions demarcate the regimes of the two metamagnetic
	transitions. The lines joining the data points are meant to be a guide to the eye.}
\label{Fig06}
\end{figure}
\end{center}
}
\subsection{Microwave Measurements}

Fig.\ref{Fig07} shows the surface impedance at 10\thinspace GHz as a function
of temperature. For a normal metal, the real and imaginary parts of the
surface impedance are equal, as is obvious from elementary electrodynamic
considerations\cite{JDJackson75}, and indeed such behaviour is observed in
the normal state for all the materials we have studied in our setup so far.
However, ${\rm DyNi_2B_2C}$ is unique in the sense that $R_s$ and
$X_s$ (where $Z_s=R_s+iX_s$) differ from each other even in the
AFM state below $T_N$. The quantity with more direct physical relevance
is the complex conductivity $\sigma_s=\sigma_1-i\sigma_2$. This
quantity was calculated from our data using the formula
$\sigma_1={\T 2\mu_0\omega R_sX_s\over\T \left(2R_sX_s\right)^2+\left(X_s^2-R_s^2\right)^2}$ and 
$\sigma_2={\T \mu_0\omega\left(X_s^2-R_s^2\right)\over\T \left(2R_sX_s\right)^2+\left(X_s^2-R_s^2\right)^2}$
and is shown in the inset of
Fig.\ref{Fig07}. The signature of the onset of the magnetic order on
$\sigma_s$ is that the imaginary part is non-zero even in the normal state.
{
\narrowtext
\begin{center}
\begin{figure}
\mbox{\epsfig{file=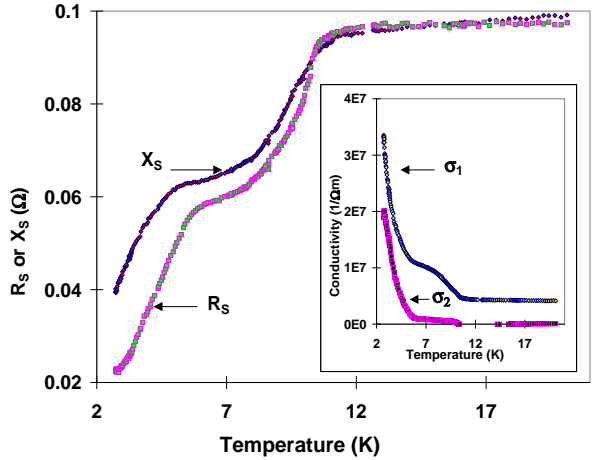, height=3.2 truein, width=2.5 truein, angle=-90}}
\caption{The real and imaginary parts of the microwave surface impedance of
	DyNi${}_2$B${}_2$C as a function of temperature. Inset : The real and
	imaginary parts of the complex conductivity against temperature.}
\label{Fig07}
\end{figure}
\end{center}
}
To see this from another perspective, we have plotted the absolute value of the surface
impedance, $Z_s(T)=\sqrt{R_s^2+X_s^2}$, and the phase angle
$\theta(T)=\tan^{-1}(X_s/R_s)$ in Fig.\ref{Fig08}. Although there is an
overall similarity of shape of $Z_s$ vs. $T$ between the microwave
measurements and the ones measured using the radio frequency technique and
the regular 4-probe DC resistivity, the microwave measurements yield
additional information associated with conductivity relaxation. For a
``conventional'' superconductor, the phase angle is expected to change
sharply from $45^{\circ}$ in the normal state tending towards $90^{\circ}$ as
$T\rightarrow 0$ below the superconducting transition temperature, and such
transitions have indeed been observed by us for most types of
superconductors. Typical data for a single crystal of ${\rm YNi_2B_2C}$
\cite{TJacobs95b} is shown in the inset of Fig.\ref{Fig08} for reference. The
phase angle starts to increase relatively sharply at the onset of the
antiferromagnetic transition and continues to increase for about
0.5\thinspace K below $T_{N}$ below which it becomes flat again until the
superconducting transition sets in. The deviation of $\theta $ from
$45^{\circ}$ in DyNi${}_2$B${}_2$C even before the superconducting
transition is a possible consequence of conductivity relaxation and is
discussed below. Below the superconducting transition at $T_c$, $\theta$
rises as $T$ is decreased as is expected for a superconductor.
{
\narrowtext
\begin{center}
\begin{figure}
\mbox{\epsfig{file=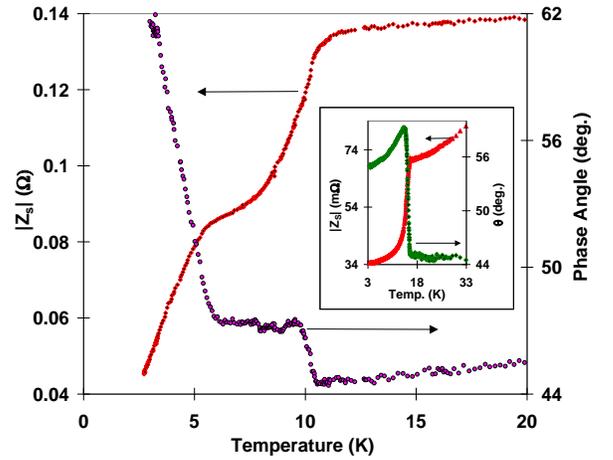, height=3.2 truein, width=2.5 truein, angle=-90}}
\caption{The magnitude and phase angle of microwave surface impedance of
	DyNi${}_2$B${}_2$C as a function of temperature. Inset : The same quantities
	measured on a single crystal of $\rm YNi_2B_2C$, from ref.\,17.}
\label{Fig08}
\end{figure}
\end{center}
}
\subsection{Conductivity relaxation in the AFM state }

The observed data can be described by a frequency dependent complex
conductivity $\sigma(\omega)=\sigma_0/(1+i\omega\tau)$ where $\sigma_0$ is
the dc conductivity. From $Z_s=\sqrt{\mu_0\omega i/\sigma(\omega)}
=(\mu_0\omega/\sigma_0)^{1/2}(i-\omega\tau)^{1/2}$, we obtain the
temperature dependence of $\omega\tau(T)={\frac{\T X_s^2-R_s^2}
{\T 2X_sR_s}}$ in the region $T_c<T<21\,$K.
The resulting $\omega\tau(T)$ is shown in the bottom panel of Fig.\ref{Fig09}. Also
shown in the same figure on the top is
$\rho_0=1/\sigma_0={\T 2R_sX_s\over\T \mu_0\omega}=m/ne^2\tau$
where $m$ is the band mass, $n$ is the density of charge carrier,
$e$ is the electronic charge and $\tau$ is the scattering life time.
Finally $\omega_p^2=\sigma_0/\tau\epsilon_0$ where
$\omega_p$ is the plasma frequency can also be calculated. For $T<T_N$,
$\omega_p\sim 10^{15}$\,Hz is comparable to plasma frequencies of other metals.
{
\narrowtext
\begin{center}
\begin{figure}
\mbox{\epsfig{file=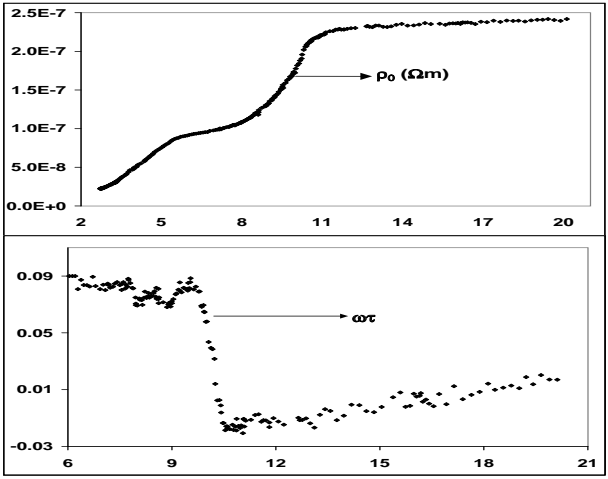, height=2.5 truein, width=3.2 truein}}
\caption{temperature dependence of some of the material parameters of $\rm DyNi_2B_2C$.
	TOP : extracted dc resistivity $\rho_0$, BOTTOM: $\omega\tau$ where $\omega$ is the
	measurement frequency and $\tau$ is the scattering life time.}
\label{Fig09}
\end{figure}
\end{center}
}
For $T>T_N$, $\omega\tau\rightarrow 0$, consistent with $R_s=X_s$ in
the normal state. This is the Hagens-Rubens limit of a Drude metal. In the AFM
state for $T<T_N$, $\omega\tau$ increases with decreasing $T$,
saturating at a value of around $0.09$. This indicates that the scattering
rate $\tau^{-1}$ approaches a value of $110\,$GHz. Thus at the measurement
frequency of $10\,$GHz we are pushing into the relaxation regime.

The origin of this conductivity relaxation is related to the development of
magnetic order, as is indicated by the temperature dependence of $\omega
\tau(T)$. The scattering of conduction electrons by spin waves in the AFM
state can give rise to the conductivity relaxation observed here.
Conductivity relaxation is also observed in other cases such as spin
density wave systems \cite{HHSJavadi85a} and heavy fermion metals \cite{SDonovan97a}.

\subsection{\protect\smallskip Superconducting state}

Since $\tau$ is more or less temperature independent for $T_N>T>T_c$
other than the sharp change in the immediate vicinity of $T_N$, we can
plausibly assume that this quantity remains independent of temperature even
below $T_c$ in the superconducting state. Since $\omega\tau\sim 0.09$
around $T_c$, the ``effective'' conductivity below $T_c$ is assumed to
be given by
$\sigma_{s,eff}=\sigma_{1,eff}-i\sigma_{2,eff}=(\sigma_1-i\sigma_2)/(1+0.09i)$.
Thus we are renormalizing the frequency-dependent conductivity in the
superconducting state from the AFM state. The resulting temperature
dependence of $\sigma_{1,eff}$ and $\sigma_{2,eff}$ in the superconducting
state are shown in Fig.\ref{Fig10}. Notice that $\sigma_{2,eff}=0$ for
$T>T_c$ which indicates that this quantity is the correct measure of the
superfluid density in the system. In the superconducting state, the
temperature dependence of $\sigma_{2,eff}$ is anomalously broad and may
indicate strong pairbreaking effects\cite{BKCho96a}. This is the fact that $\sigma_{2,eff}$
continues to rise even for $T<T_c/2$, atypical of a ``conventional''
superconductor. The origin of this is the very broad normal to
superconducting transition seen in this material which is also observed from
dc resistivity measurements\cite{BKCho95a,CVTomy95a}. This may be related to
the strong scattering we have described in earlier in the AFM state. Thus
although antiferromagnetism does not inhibit the formation of the
superconducting state, it appears to have a strong effect on the
electrodynamic properties. 
{
\narrowtext
\begin{center}
\begin{figure}
\mbox{\epsfig{file=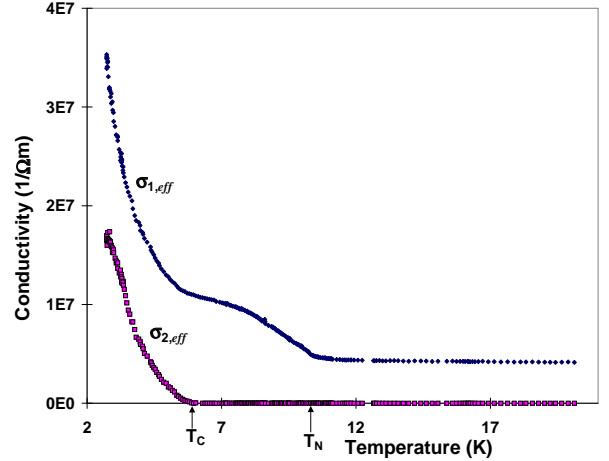, height=3.2 truein, width=2.5 truein, angle=-90}}
\caption{The ``effective'' values of $\sigma_1$ and $\sigma_2$.}
\label{Fig10}
\end{figure}
\end{center}
}
\section{Conclusion}

In summary, a comprehensive set of measurements of the electrodynamic
properties of ${\rm DyNi_2B_2C}$ in the radio frequency and microwave
regimes have been carried out, revealing some very interesting effects of
magnetic and superconducting order on transport properties. The presence of
magnetic Dy ions strongly affects the transport properties. In the AFM
state, establishment of long range magnetic order leads to a strong
reduction in scattering. Establishment of antiferromagnetic order gives rise
to anomalous increase of electron scattering time which makes the real and
imaginary parts of complex microwave surface impedance differ from each
other. We have interpreted this behaviour in terms of conductivity relaxation
in the AFM state.

Antiferromagnetism also has a distinctive signature on the radio frequency
skin depth measurements. The angular dependence of this quantity suggests
anisotropic vortex response. Two field induced metamagnetic transitions were
observed in the antiferromagnetic state, at all temperatures below $T_{N}$.
A theoretical framework taking into account the scattering of the conduction
electrons from the 2-D sheets of ferromagnetically coupled rare earth
moments is likely to be required for a clear understanding of all the
observed effects.

While several interesting results have emerged from the present
measurements, the results also suggest new avenues to explore in further
work and also call for a better theoretical understanding of these novel
materials.

\section{Acknowledgments}

Stimulating discussions with Balam A. Willemsen, S.~Oxx, Benjamin
Revcolevschi and T.~Jacobs are thankfully acknowledged. This research was
supported by the National Science Foundation through grant No. 9623720.

\end{multicols}
\bibliographystyle{prsty}
\bibliography{/home/durga/u/durga/papers/strings, /home/durga/u/durga/papers/big}

\begin{thebibliography}{10}

\bibitem{RNagarajan94a}
R. Nagarajan {\it et~al.}, Phys. Rev. Lett. {\bf 72},  274  (1994).

\bibitem{RJCava94a}
R.~J. Cava {\it et~al.}, Nature {\bf 367},  146  (1994).

\bibitem{KDDRathnayaka96a}
K. Rathnayaka, D. Naugle, B. Cho, and P. Canfield, Phys. Rev. B {\bf 53},  5688
   (1996).

\bibitem{YDeWilde97a}
Y.~D. Wilde {\it et~al.}, Phys. Rev. Lett. {\bf 78},  4273  (1997).

\bibitem{RJCava94b}
R.~J. Cava {\it et~al.}, Nature {\bf 367},  252  (1994).

\bibitem{HEisaki94a}
H. Eisaki {\it et~al.}, Phys. Rev. B {\bf 50},  647  (1994).

\bibitem{BKCho95a}
B.~K. Cho, P.~C. Canfield, and D.~C. Johnston, Phys. Rev. B {\bf 52},  R3844
  (1995).

\bibitem{MEMassalami95a}
M.~E. Massalami, S.~L. Bu{\'d}ko, B. Giordanengo, and E.~M. Baggio-Saitovitch,
  Physica C {\bf 244},  41  (1995).

\bibitem{CVTomy95b}
C.~V. Tomy, G. Balakrishnan, and D.~M. Paul, Physica C {\bf 248},  349  (1995).

\bibitem{HCKu80a}
H.~C. Ku, F. Acker, and B.~T. Matthias, Phys. Lett. A {\bf 76A},  399  (1980).

\bibitem{FGAliev94a}
F.~G. Aliev {\it et~al.}, Physica B {\bf 194-196},  171  (1994).

\bibitem{DHWu90a}
D.-H. Wu and S. Sridhar, Phys. Rev. Lett. {\bf 65},  2074  (1990).

\bibitem{SOxx96a}
S. Oxx {\it et~al.}, Physica C {\bf 264},  103  (1996).

\bibitem{SSridhar96a}
S. Sridhar {\it et~al.}, Phys. Rev. Lett. {\bf 77},  2145  (1996).

\bibitem{MREskildsen97a}
M.~R. Eskildsen {\it et~al.}, Phys. Rev. Lett. {\bf 78},  1968  (1997).

\bibitem{SSridhar88b}
S. Sridhar and W.~L. Kennedy, Rev. Sci. Instrum. {\bf 59},  531  (1988).

\bibitem{TJacobs95b}
T. Jacobs {\it et~al.}, Phys. Rev. B {\bf 52},  R7022  (1995).

\bibitem{TJacobs95d}
T. Jacobs {\it et~al.}, Phys. Rev. Lett. {\bf 75},  4516  (1995).

\bibitem{HSrikanth97a}
H. Srikanth {\it et~al.}, Phys. Rev. B {\bf 55},  R14733  (1997).

\bibitem{MXu94a}
M. Xu {\it et~al.}, Physica C {\bf 227},  321  (1994).

\bibitem{PDervenagas95a}
P. Dervenagas {\it et~al.}, Physica B {\bf 212},  1  (1995).

\bibitem{MSLin95a}
M.~S. Lin {\it et~al.}, Physica C {\bf 249},  403  (1995).

\bibitem{JWLynn96b}
J.~W. Lynn {\it et~al.}, Physica B {\bf 223 \& 224},  66  (1996).

\bibitem{PCCanfield97b}
P.~C. Canfield and S.~L. Bud'ko, Journal of Alloys and Compounds  , in Press.

\bibitem{SOxx97a}
S. Oxx {\it et~al.},   , to be Published.

\bibitem{PCCanfield94a}
P.~C. Canfield {\it et~al.}, Physica C {\bf 230},  397  (1994).

\bibitem{PCCanfield97a}
P.~C. Canfield {\it et~al.}, Phys. Rev. B {\bf 55},  970  (1997).

\bibitem{AIBuzdin86a}
A.~I. Buzdin and L.~H. Bulaevsk\u{ii}, Sov. Phys. Usp. {\bf 29},  412  (1986).

\bibitem{MBMaple95a}
M.~B. Maple, Physica B {\bf 215},  110  (1995).

\bibitem{RKonno95a}
R. Konno, Physica B {\bf 206 \& 207},  638  (1995).

\bibitem{MBMaple82}
M.~B. Maple and {\O}. Fischer, {\em Superconductivity in Ternary Compounds II :
  Superconductivity and Magnetism} (Springer-Verlag, New York, 1982), "Topics
  in Current Physics Vol. 34".

\bibitem{MLKulic95a}
M.~L. Kuli{\'c}, A.~I. Lichenstein, E. Goreatchkovski, and M. Mehring, Physica
  C {\bf 244},  185  (1995).

\bibitem{HIwasaki86a}
H. Iwasaki, M. Ikebe, and Y. Muto, Phys. Rev. B {\bf 33},  4669  (1986).

\bibitem{AIBuzdin90a}
A.~I. Buzdin, S.~S. Krotov, and D.~A. Kuptsov, Sol. St. Comm. {\bf 75},  229
  (1990).

\bibitem{OWong89a}
O. Wong, H. Umezawa, and J.~P. Whitehead, Physica C {\bf 158},  32  (1989).

\bibitem{IRFisher96a}
I.~R. {Fisher \it et. al}, J. Low. Temp. Phys. {\bf 105},  1623  (1996).

\bibitem{IRFisher97a}
I.~R. Fisher, J.~R. Cooper, and P.~C. Canfield, Phys. Rev. B  , to appear as a
  brief report.

\bibitem{ZHossain96a}
Z. Hossain {\it et~al.}, Physica B {\bf 223 \& 224},  99  (1996).

\bibitem{JDJackson75}
J.~D. Jackson, {\em Classical Electrodynamics}, 2 ed. (Wiley, New York, 1975).

\bibitem{HHSJavadi85a}
H.~H.~S. Javadi {\it et~al.}, Phys. Rev. Lett. {\bf 55},  1216  (1985).

\bibitem{SDonovan97a}
S. Donovan, A. Schwartz, and G. Gr{\"u}ner, Phys. Rev. Lett. {\bf 79},  1401
  (1997).

\bibitem{BKCho96a}
B.~K. Cho, P.~C. Canfield, and D.~C. Johnston, Phys. Rev. Lett. {\bf 77},  163
  (1996).

\bibitem{CVTomy95a}
C.~V. Tomy {\it et~al.}, Phys. Rev. B {\bf 52},  9186  (1995).

\end{thebibliography}
\end{document}